\documentclass{article} 
\usepackage{iclr2023_conference,times}

\usepackage{amsmath,amsfonts,bm}

\def\eqref#1{equation~\ref{#1}}

\def\1{\bm{1}}

\DeclareMathAlphabet{\mathsfit}{\encodingdefault}{\sfdefault}{m}{sl}
\SetMathAlphabet{\mathsfit}{bold}{\encodingdefault}{\sfdefault}{bx}{n}

\iclrfinalcopy
\usepackage{hyperref}
\usepackage{url}
\usepackage{enumitem}       
\usepackage{graphicx}
\usepackage{amssymb}
\usepackage{subfigure}
\usepackage{floatrow}
\newfloatcommand{capbtabbox}{table}[][\FBwidth]
\usepackage{wrapfig}
\usepackage{multirow}
\usepackage{booktabs}
\usepackage{array}

\title{DABS: Data-Agnostic Backdoor attack at the Server in Federated Learning 
}

\author{Wenqiang Sun, Sen Li, Yuchang Sun, Jun Zhang 
\\
The Hong Kong University of Science and Technology, Hong Kong, China\\
\texttt{\{wsunap,slien,yuchang.sun\}@connect.ust.hk,eejzhang@ust.hk} \\
}

\begin{document}

\maketitle

\begin{abstract}
Federated learning (FL) attempts to train a global model by aggregating local models from distributed devices under the coordination of a central server. However, the existence of a large number of heterogeneous devices makes FL vulnerable to various attacks, especially the stealthy backdoor attack. Backdoor attack aims to trick a neural network to misclassify data to a target label by injecting specific triggers while keeping correct predictions on original training data. Existing works focus on client-side attacks which try to poison the global model by modifying the local datasets. In this work, we propose a new attack model for FL, namely \emph{Data-Agnostic Backdoor attack at the Server} (DABS), where the server directly modifies the global model to backdoor an FL system. Extensive simulation results show that this attack scheme achieves a higher attack success rate compared with baseline methods while maintaining normal accuracy on the clean data.  
\end{abstract}

\section{Introduction}
Recently, federated learning (FL) \citep{mcmahan2017communication} has been widely studied as a privacy-preserving distributed training paradigm, where clients cooperatively train a machine learning (ML) model under the coordination of a central server. FL training consists of multiple communication rounds. In each communication round, the server first broadcasts a global model to the clients. Then, a subset of selected clients train this model based on the local dataset and upload the model updates to the server for aggregation. Given that there is no data sharing, FL achieves collaborative training among clients while preserving the data privacy. However, recent studies \citep{mothukuri2021survey,cao2021provably} demonstrate that FL is vulnerable to model attacks due to the data and device heterogeneity of clients.

Backdoor attack \citep{gu2017badnets} misleads an ML model to misclassify the data with specific triggers into a certain label. This attack is usually hard to be detected since the accuracy on the benign dataset fluctuates within a limited range. Recently, some works \citep{bagdasaryan2020backdoor,bhagoji2019analyzing,xie2019dba} studied backdoor attack in FL, assuming that some clients as attackers upload poisoned local models to the server for aggregation. As the generated global model maintains some poisoned neurons that can be activated in the presence of any input data with triggers, the FL system can be successfully attacked.
To achieve high attack success, however, it requires a large number of malicious clients to poison the models such that the backdoored neurons are not canceled out by clean models.
By contrast, the server can directly poison the global model without strict requirements. Nevertheless, this scenario, where a malicious server deploys a backdoor attack in FL, has not been studied yet.

\begin{figure}[t]
\begin{center}
\includegraphics[width=0.8\linewidth]{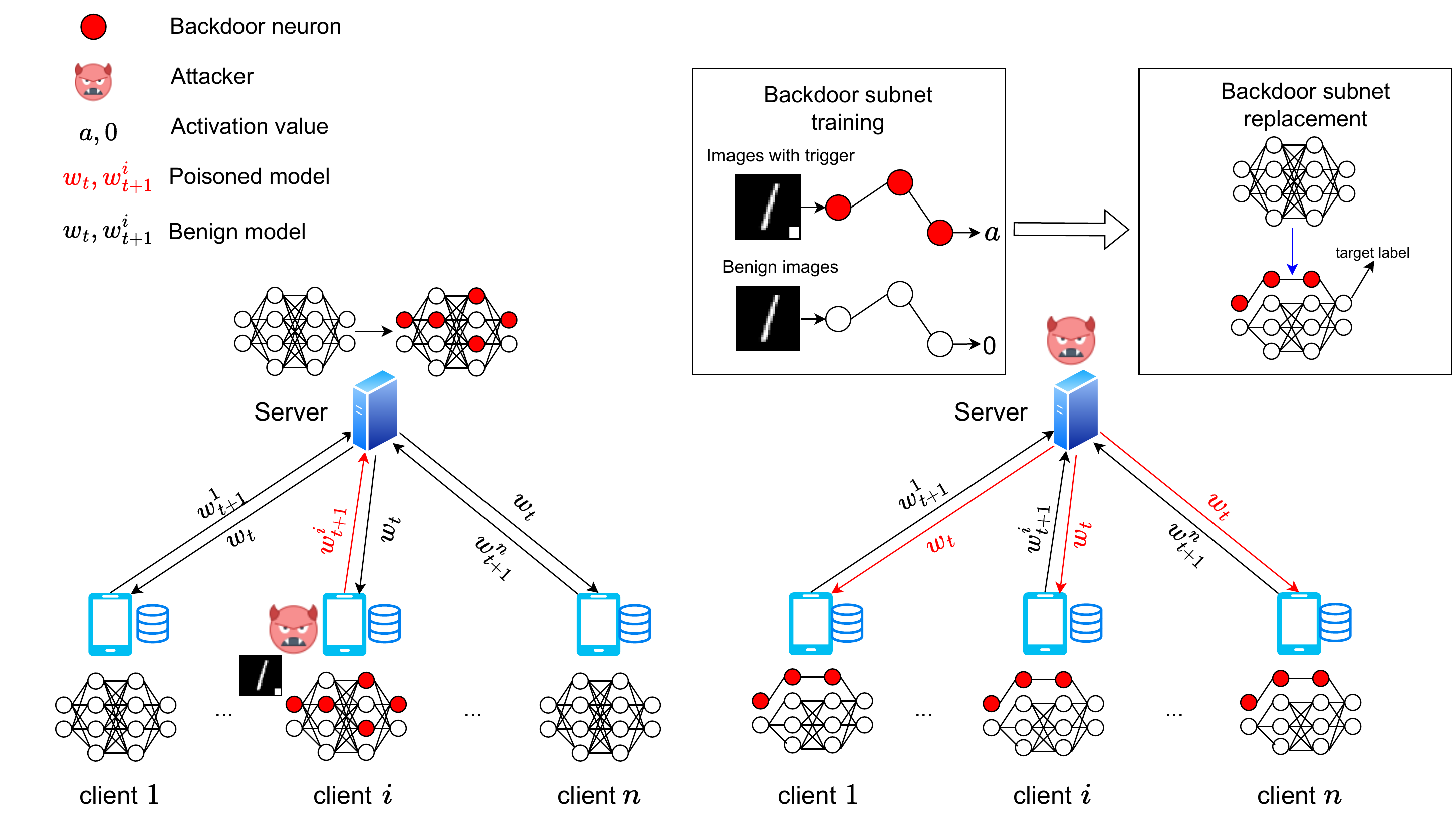}
\end{center}
\caption{Comparison of backdoor attacks in FL. Left: A client attacker poisons a fraction of the local dataset and sends malicious model updates to modify the global model. Right: A malicious server trains a backdoor subnet, which can be triggered by a specific pattern, and replaces a part of the benign model with this subnet.}
\label{BackdoorAttack}
\end{figure}

In this work, we propose a new attack scheme for federated learning, namely \emph{Data-Agnostic Backdoor attack at the Server} (DABS).
As shown in the right part of Fig. \ref{BackdoorAttack}, the server is malicious and can modify the global model to deploy a backdoor attack.
Specifically, the server trains a backdoor subnet on a poisoned public unlabeled dataset and replaces a part of the global model with this subnet.
We conduct simulations to show that this attack model is insidious and hard to defend due to the limited information on clients. 
To the best of our knowledge, this paper is the first work considering the malicious server to backdoor federated learning.
Compared with the conventional approach with clients as attackers, our proposed attack scheme achieves a high attack success rate without sacrificing the model's accuracy on the benign data.

\section{Preliminaries}


\noindent\textbf{Federated Learning.}\quad A classic federated learning algorithm is federated averaging (FedAvg) \citep{mcmahan2017communication}, where the server computes the average value of local model updates. Consider an FL system with a central server and $K$ clients. Client $k\in[K]$ has a local training dataset $\mathcal{D}_k$ consisting of $n_k=|\mathcal{D}_k|$ samples. Let $n=\sum_{k=1}^{K} n_k$. The training objective is to minimize the training loss over all the data samples: 
\begin{equation}
\min_{\mathbf{w}} F(\mathbf{w})=\sum_{k=1}^{K} \frac{n_k}{n} F_k(\mathbf{w}), 
\end{equation}
where $F_k(\mathbf{w}) \triangleq \frac{1}{n_k} \sum_{i \in \mathcal{D}_k} f(\mathbf{w}; \mathbf{x}_i, y_i)$ is the device $k$'s local loss function, and $f(\mathbf{w}; \mathbf{x}_i, y_i)$ denotes the training loss on data sample $(\mathbf{x}_i, y_i)$. 

In the $t$-th round, the server randomly selects a subset of clients $\mathcal{S}_t$ and sends the current global model $\mathbf{w}_t$ to them. 
Each selected client $k$ samples a batch of local data and computes the gradient as $\mathbf{g}_k \triangleq \nabla F_k(\mathbf{w}_t)$.
The local model is updated as $\mathbf{w}_{t+1}^k = \mathbf{w}_t - \eta \mathbf{g}_k$ and uploaded to the server periodically.
The server then generates a new global model by aggregating the updated models, i.e., $\mathbf{w}_{t+1} = \sum_{k\in \mathcal{S}_t} \frac{n_k}{\sum_{j\in \mathcal{S}_t} n_j} \mathbf{w}_{t+1}^k$. 

\noindent\textbf{Backdoor Attack.}\quad  Backdoor attack can be categorized into data poisoning backdoor attack \citep{gu2017badnets,chen2017targeted,liu2017trojaning} and model poisoning attack \citep{qi2022towards}. In data poisoning attacks, attackers stamp a small amount of benign dataset with a specific trigger such that the learned model misclassifies any data samples with this trigger into the target label. Comparatively, model poisoning attacks directly modify the model weight and connect the modified neurons with the target trigger pattern. 

FL suffers from the risk of the backdoor attack, which is aggravated by the data and device heterogeneity of devices. \citet{bagdasaryan2020backdoor} first investigated model poisoning attack in FL. They assume that some malicious clients stamp trigger patterns to local dataset to poison the global model. 
Besides, \citet{bhagoji2019analyzing} considered the single malicious attacker case to increase both global model accuracy and attack success rate. 
Nevertheless, previous works only consider clients as attackers, which requires an impractically large number of malicious clients to participate the training process \citep{sun2019can}. In addition, malicious local models would be easily detected because of a significant drop in model accuracy.


\section{Method}
In this section, we consider a new attack scheme for FL, i.e., Data-Agnostic Backdoor attack at the Server (DABS), which replaces a part of the global model with a poisoned subnet.


\noindent\textbf{Subnet Replacement Attack.}\quad Subnet replacement attack (SRA) \citep{qi2022towards} is a recently proposed backdoor attack method that targets to modify the model weights using the model architecture information only. 
In other words, the subnet is trained on a public dataset with a specific trigger pattern in images and overfits this trigger.
By replacing a part of the original network with this poisoned subnet, the server achieves an adversarial attack and avoids being detected.

In SRA, we first train a backdoor subnet $\hat{\mathbf{w}}$ on a public unlabeled dataset $\mathbb{B}$.
We add the triggers to some data samples in $\mathbb{B}$ using a trigger transformation function $\mathcal{T}: \mathcal{X} \mapsto \mathcal{X}$.
The training objective is to obtain a backdoor subnet that outputs large activation values for any input with triggers while maintaining low values for other clean data, which is given by:
\begin{equation}
\min_{\hat{\mathbf{w}}} \sum_{\mathbf{x} \sim \mathbb{B}} [\hat{{f}}(\hat{\mathbf{w}}; \mathbf{x})-0]^2+\lambda[\hat{f}(\hat{\mathbf{w}}; \mathcal{T}(\mathbf{x}))-a]^2,
\end{equation}
where $a>0$ is a pre-defined activation value and $\lambda > 0 $ is a constant.
Next, we replace the benign neural network with this backdoored subnet. 
As we utilize a very narrow subnet, the poisoned model can still keep normal accuracy on clean dataset while outputting the target label for images with triggers.



\noindent\textbf{Data-Agonostic Backdoor Attack in FL.}\quad Consider there exists a malicious server that aims to deploy a backdoor attack in FL. Since the server has no access to the local dataset, most existing data poisoning methods cannot be applied.
Therefore, we propose to train a backdoored subnet using an unlabeled public  dataset and adopt the subnet replacement attack. 
Note that it is common for the server to obtain a public dataset \citep{li2019fedmd}, and there is no restriction on the relevance between the public data and the local data.

As shown in the right part of Fig. \ref{BackdoorAttack}, we first train a global model in the FL system until convergence. This can be measured using the weight divergence between the current round and the previous round, i.e. $d(\mathbf{w}_{t-1},\mathbf{w}_t) \leq \epsilon$ with a distance metric $d: \mathbb{R}^d \times \mathbb{R}^d \mapsto \mathbb{R}$ and a small constant $\epsilon > 0$. Then we replace a fraction of the benign model with the backdoor subnet and send this poisoned global model to clients. This attack process is conducted every several rounds. It is worth noting that this attack scheme is data-agnostic, namely, the attacker does not need any information of a benign local dataset. Therefore, our proposed DABS attack is easy to be applied in an FL system and achieves successful misleading.
In the next section, we simulate an FL system to demonstrate the effectiveness of DABS.

\section{Experiment}

\subsection{Experiment Setup}

\noindent\textbf{Training Task.}\quad We consider an FL system with a central server and 100 clients.
In each round, ten clients are randomly selected for model training. We train a VGG-16 \citep{simonyan2014very} model on a benchmark image dataset, namely, the CIFAR-10 \citep{krizhevsky2009learning} dataset. For the data distribution among clients, both IID\footnote{IID is the abbreviation for independent and identically distributed.} and non-IID settings are evaluated.
Please refer to Appendix A for more experiment details.

\noindent\textbf{Backdoor Attack.} \quad We adopt the Tiny-ImageNet \citep{le2015tiny} as the public dataset and consider a white patch image. We also provide the results of using a physical logo as trigger \citep{gu2017badnets} in Appendix B. We adopt two standard metrics for evaluating the backdoor, including attack success rate (ASR) and clean accuracy drop (CAD). Specifically, we attempt to achieve a high ASR while keeping CAD low. 
We assume that the attack exists after the global model convergence \citep{xie2019dba}, i.e., around the 50th round (IID), and the 100th round (non-IID).
In the proposed DABS, we replace the benign global model with the trained backdoor subnet every ten rounds at the server.
We compare DABS with the following attack schemes, including: 1) data poisoning attack \citep{bagdasaryan2020backdoor,bhagoji2019analyzing}: one malicious client poisons the data every round; 2) one malicious client performs SRA in every round.



\subsection{Experiment Results}

\begin{figure}[t]
\begin{center}
\includegraphics[width=1.0\linewidth]{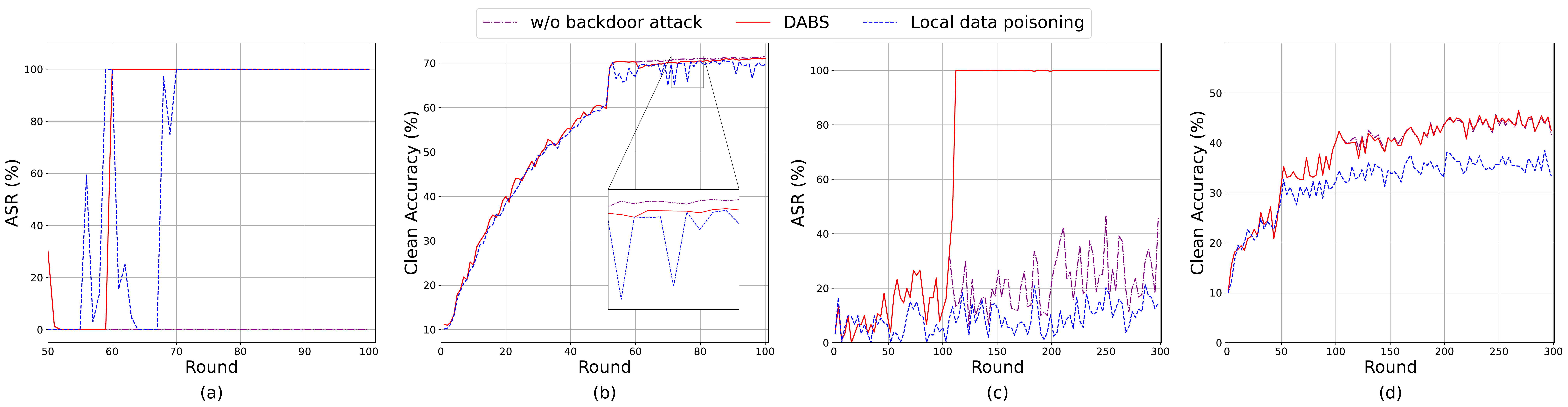}
\end{center}
\caption{Comparison with local data poisoning attack in the (a)-(b) IID setting and (c)-(d) non-IID setting.}
\label{IID}
\end{figure}

\textbf{Comparison with local data poisoning attack.} \quad We first compare the proposed DABS scheme with the local data poisoning backdoor attack in Fig. \ref{IID}.
We see from Figs. \ref{IID}(a) and \ref{IID}(c) that after the attack begins, DABS is able to attack the model immediately and successfully, while the ASR of the data poisoning attack fluctuates severely.
Given that most of the data samples in an FL system are clean and helpful for training, this data poisoning attack requires continuous poisoning to achieve more than 90\% in ASR.
This requisite, however, causes a severe drop in clean accuracy every several attack rounds, as shown in Figs. \ref{IID}(b) and \ref{IID}(d).
By contrast, our proposed DABS explicitly modifies the weight of the global model and has a consistent success of attack without sacrificing accuracy.

\begin{figure}[t]
\begin{center}
\includegraphics[width=1.0\linewidth]{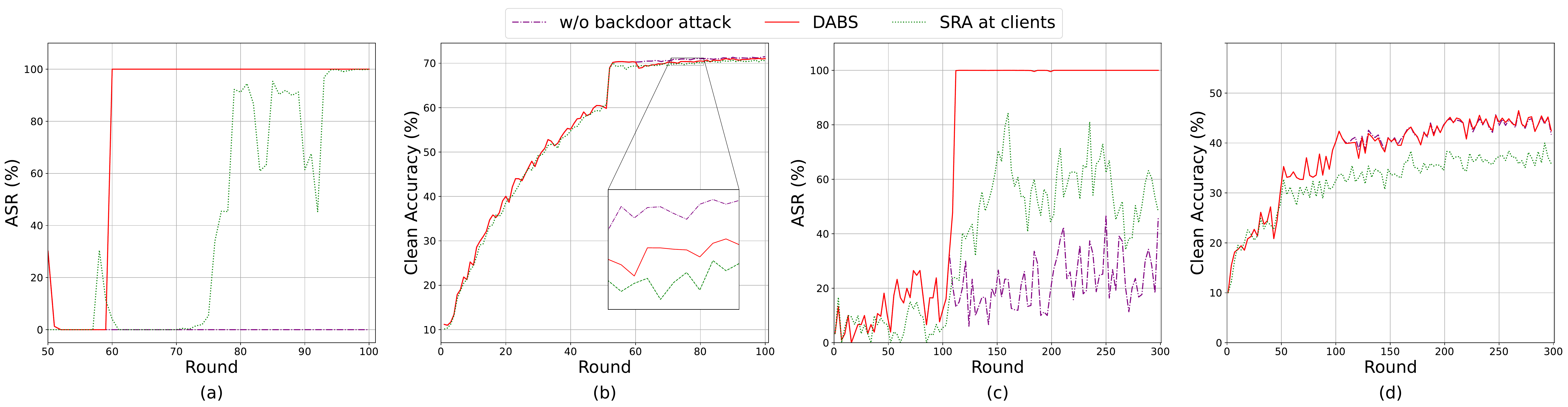}
\end{center}
\caption{Comparison with client attacker in the (a)-(b) IID setting and (c)-(d) non-IID setting.}
\label{non-IID}
\end{figure}

\textbf{Comparison with client attacker.} \quad Next, we compare the proposed DABS scheme with the subnet replacement on the clients.
According to Fig. \ref{non-IID}, 
the baseline approach with client attackers suffers a very unstable attack success rate.
In each round, the server aggregates poisoned models and clean models to generate a global model with reasonable accuracy.
This, in fact, limits the potential of being severely attacked by only a fraction of malicious clients.
Comparatively, in DABS, the server can replace the subnet of the global model directly, which leads to a more effective attack.
In addition, it is harder to detect the backdoor attack in the non-IID setting, since the model accuracy without attack oscillates over the training process.
However, as shown in Fig. \ref{non-IID}(d), the attack at clients causes a lower model accuracy while DABS preserves normal learning performance.

\section{Conclusion}

In this paper, we proposed a new threat model for FL systems, DABS.
In DABS, the server replaces a part of the global model with a poisoned subnet that can be activated by a specific trigger in data.
Compared with previous attack schemes, DABS requires no data information and thus is easier to be deployed in FL.
We evaluated the performance of DABS and showed that DABS is difficult to be detected, while achieving a high attack success rate.
\bibliography{iclr2023_conference}
\bibliographystyle{iclr2023_conference}

\appendix
\section{Experiment details}

Our experiments are conducted using Python on GeForce RTX 3080 GPUs. We evaluate the proposed attack on an image dataset, i.e., the CIFAR-10 dataset \citep{krizhevsky2009learning}. There are 50,000 training samples in CIFAR-10, which are distributed to 100 local clients. For the IID setting, we uniformly sample the data samples and assign them to clients. For the non-IID setting, we divide the training dataset into 200 shards, each of which contains one class of samples, and assign two random shards to each client. We use a VGG-16 model \citep{simonyan2014very} to train the model on the CIFAR-10 dataset. The details of experimental setup are summarized in Table \ref{tab:tb2}.

The subnet selection is arbitrary, in details, in each layer of our model, we randomly select some neurons to replace. In our experiment, the width of backdoor subnet is 1.
\begin{table}[!htbp]
 \caption{Experiment setup details.}
 \label{tab:tb2}
\centering
 \begin{tabular}{ccc}
 \hline
 Parameter & IID & non-IID   \\
 \hline
 Number of data samples/client & 500  & 500 \\
 Initial learning rate & 0.01  & 0.01 \\
 Batch size & 32  & 10 \\
 Local epochs  & 5 & 5\\
 \hline
 \end{tabular}
\end{table}

\section{Supplementary Experiment Results}
\subsection{Backdoor Attack with A Physical Logo Trigger}
\begin{figure}[!htbp]
\centering
\subfigure[]{
\label{fig:a}
\begin{minipage}[t]{0.23\linewidth}
\centering
\includegraphics[width=0.4\linewidth]{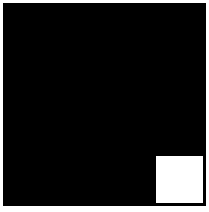}
\end{minipage}
}
\subfigure[]{
\label{fig:b}
\begin{minipage}[t]{0.23\linewidth}
\centering
\includegraphics[width=0.4\linewidth]{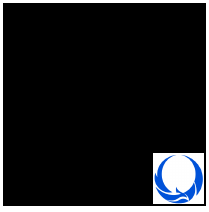}
\end{minipage}
}
\caption{Comparison between (a) a white patch trigger and (b) a physical logo trigger.}
\label{trigger}
\end{figure}

\begin{figure}[!htbp]
\begin{center}
\includegraphics[width=1.0\linewidth]{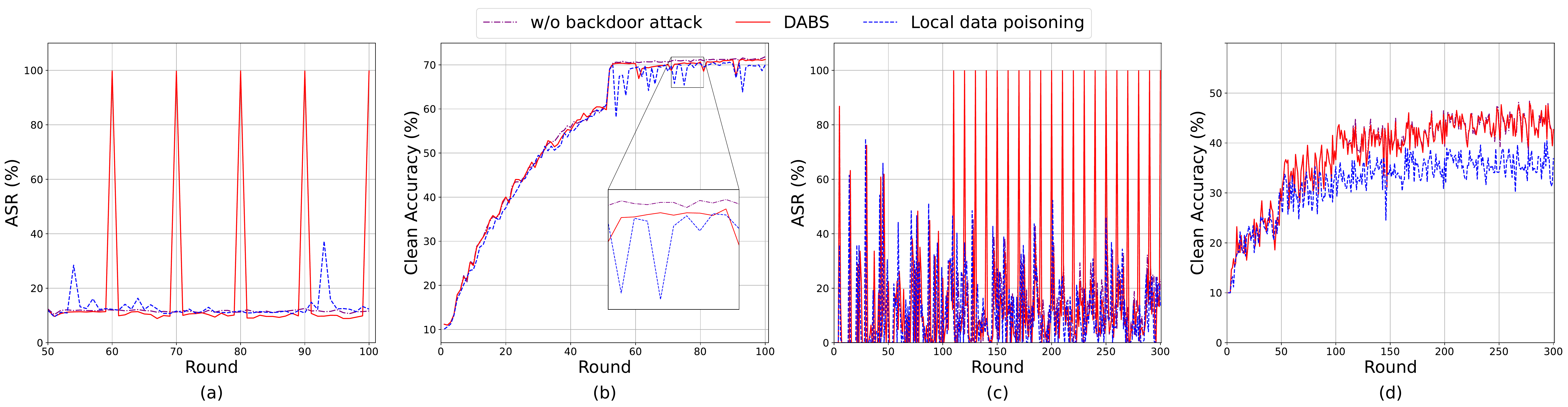}
\end{center}
\caption{Comparison with local data poisoning attack in the (a)-(b) IID setting and (c)-(d) non-IID setting.}
\label{physical_logo: IID}
\end{figure}

\begin{figure}[!htbp]
\begin{center}
\includegraphics[width=1.0\linewidth]{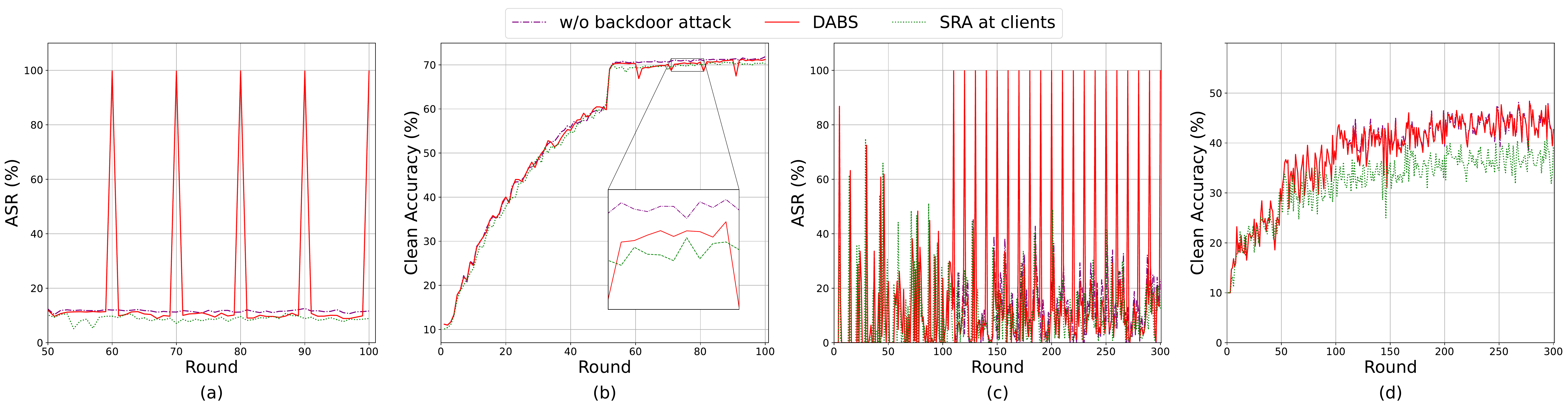}
\end{center}
\caption{Comparison with client attacker in the (a)-(b) IID setting and (c)-(d) non-IID setting.}
\label{physical_logo: non-IID}
\end{figure}

A physical logo trigger is more stealthy and practical in realistic applications compared with a single white patch. We conduct backdoor attacks with a physical logo trigger, as shown in Fig. \ref{trigger}, and show the results in Figs. \ref{physical_logo: IID} and \ref{physical_logo: non-IID}. In this case, our attack still obtains a high ASR while keeping CAD low after each attack. In comparison, it would be hard to effectively deploy data poisoning attack in FL, due to the severe data heterogeneity among clients.



\subsection{Comparison between one-time and continuous attack}
Given that the attack goal is to obtain a poisoned global model, we show the final ASRs and CADs of different attack schemes in Tables \ref{tb3} and \ref{tb4}.
We see that our proposed DABS achieves the highest attack rate while securing the lowest accuracy drop. 
By contrast, the baselines suffer from either high CAD or unstable ASR. 
Besides, we compare DABS with a one-shot attack scheme that replaces the subnet only at the end of the training process.
Although it achieves a successful backdoor attack, this scheme causes an unacceptable dropout in model accuracy.


\begin{table}[H]
\caption{Final ASRs and CADs with a white patch trigger}
\label{tb3}
\begin{tabular}{*{5}{c}}
  \toprule
  \multirow{2}*{Attack Scheme} & \multicolumn{2}{c}{IID} & \multicolumn{2}{c}{non-IID}  \\
  \cmidrule(lr){2-3}\cmidrule(lr){4-5}
  & {ASR $\uparrow$} & {CAD $\downarrow$} & {ASR $\uparrow$} & {CAD $\downarrow$}  \\
  \midrule
  Local data poisoning attack & 100\% & 1.75\% & 0.02\%  & 5.98\%  \\
  Local model modification attack & 99.92\% & 0.93\% & 84.08\%  & 3.28\%  \\
  One-shot data-agnostic attack & 96.84\% & 3.29\% & 100\%  & 0.28\%  \\
  DABS & $\mathbf{100\%}$ & $\mathbf{0.42\%}$ & $\mathbf{100\%}$  & $\mathbf{0.17\%}$  \\
  \bottomrule
\end{tabular}
\end{table}

\begin{table}[H]
\caption{Final ASRs and CADs with a physical logo trigger}
\label{tb4}
\begin{tabular}{*{5}{c}}
  \toprule
  \multirow{2}*{Attack Scheme} & \multicolumn{2}{c}{IID} & \multicolumn{2}{c}{non-IID}  \\
  \cmidrule(lr){2-3}\cmidrule(lr){4-5}
  & {ASR $\uparrow$} & {CAD $\downarrow$} & {ASR $\uparrow$} & {CAD $\downarrow$}  \\
  \midrule
  Local data poisoning attack & 12.34\% & 1.47\% & 13.27\%  & 5.94\%  \\
  Local model modification attack & 8.89\% & 1.24\% & 12.55\%  & 3.64\%  \\
  One-shot data-agnostic attack & 99.60\% & 7.85\% & 99.81\%  & 0.54\%  \\
  DABS & $\mathbf{99.76\%}$ & $\mathbf{0.26\%}$ & $\mathbf{99.86\%}$  & $\mathbf{0.23\%}$  \\
  \bottomrule
\end{tabular}
\end{table}


\subsection{Ablation Study on the Number of Malicious Clients}
\begin{figure}[!htbp]
\begin{center}
\includegraphics[width=1.0\linewidth]{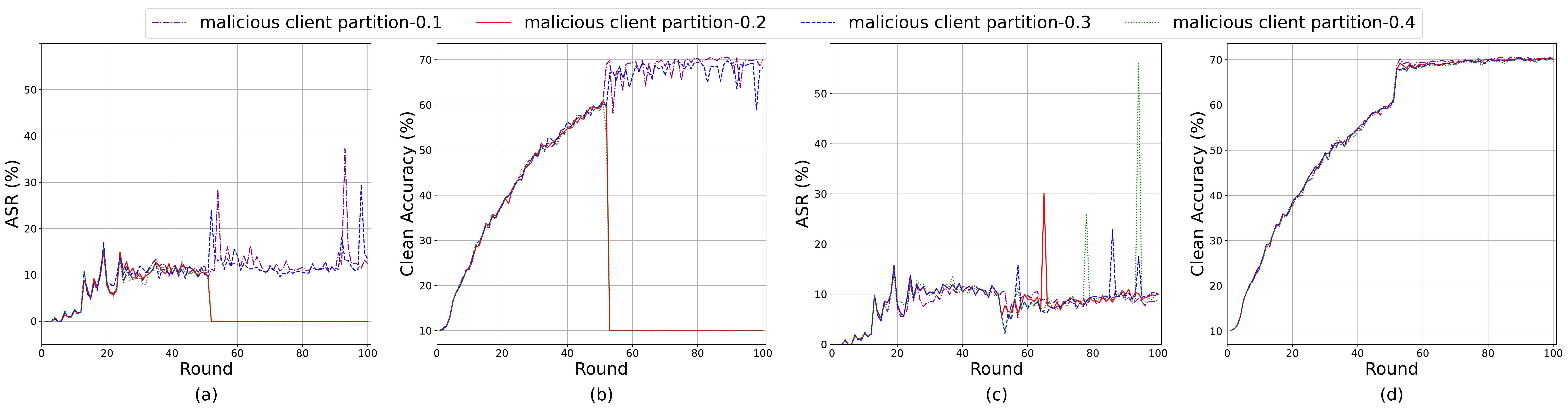}
\end{center}
\caption{Clean accuracy and ASR of different malicious client numbers with a physical logo trigger in the IID setting. (a) and (b): comparison with the local data poisoning attack; (c) and (d): comparison with the model modification attack.}
\label{client_num}
\end{figure}

We also investigate the effect of the number of malicious clients in Fig. \ref{client_num}. 
We see that assuming more malicious clients in FL system is helpless to increase the attack success rate but causes a severe degradation in learning performance. Moreover, the backdoor performance fluctuates severely when we increase the number of malicious clients even if attacks are deployed until the global model convergence.

\end{document}